\documentclass[twocolumn, twoside]{IEEEtran}

\usepackage{cite}      
\usepackage{subfigure} 
\usepackage{array}
\usepackage{multicol}

\usepackage{amsmath}   
\usepackage{xfrac}

\usepackage{authblk}
\usepackage{caption}
\usepackage{balance}

\usepackage{graphicx}
\usepackage{amssymb}
\usepackage{tabularx}

\usepackage{booktabs}

\newtheorem{theorem}{Theorem}[section]
\newtheorem{lemma}[theorem]{Lemma}

\newenvironment{proof}[1][Proof:]{\begin{trivlist}
\item[\hskip \labelsep {\bfseries #1}]}{\end{trivlist}}
\newenvironment{definition}[1][Definition]{\begin{trivlist}
\item[\hskip \labelsep {\bfseries #1}]}{\end{trivlist}}
\newenvironment{example}[1][Example]{\begin{trivlist}
\item[\hskip \labelsep {\bfseries #1}]}{\end{trivlist}}

\newcommand{\qed}{\nobreak \ifvmode \relax \else
      \ifdim\lastskip<1.5em \hskip-\lastskip
      \hskip1.5em plus0em minus0.5em \fi \nobreak
      \vrule height0.75em width0.5em depth0.25em\fi}

\title{Minimum Delay Huffman Code in Backward Decoding
Procedure}%
\author{Alireza Poostindouz and Adel Aghajan}
\affil{Department of Electrical and Computer Engineering,\\Isfahan University of Technology, Isfahan, Iran, 84156-83111.}%
\affil{\small{Email: poostindouz@ieee.org; a.aghajanabdollah@ec.iut.ac.ir}}

\graphicspath{{figs/}}

\begin{document}

\twocolumn[
  \begin{@twocolumnfalse}
    \maketitle
    \begin{abstract}
      For some applications where the speed of decoding and the fault tolerance are important, like in video storing, one of the successful answers is Fix-Free Codes. These codes have been applied in some standards like H.263+ and MPEG-4. The cost of using fix-free codes is to increase the redundancy of the code which means the increase in the amount of bits we need to represent any peace of information. Thus we investigated the use of Huffman Codes with low and negligible backward decoding delay. We showed that for almost all cases there is always a Minimum Delay Huffman Code for a given length vector. The average delay of this code for anti-uniform sources is calculated, that is in agreement with the simulations, and it is shown that this delay is one bit for large alphabet sources. Also an algorithm is proposed which will find the minimum delay code with a good performance. \\ \\
      \textbf{\textit{\small Keywords - Bidirectional Decoding Algorithm; Huffman Code; Decoding Delay; Anti-uniform Sources.}\\}
    \end{abstract}
  \end{@twocolumnfalse}

]
%

\section{Motivation}	        

Fix-free codes have larger redundancy than the Huffman codes. The bounds of the optimal fix-free codes have been investigated \cite{Savari2012}, and it is shown that the redundancy of optimal fix-free code is at most 0.8 bits greater than that of the Huffman code \cite{Khosravifard2012}. However, still it is worth finding a Huffman code with negligible backward decoding delay, because in the applications where bidirectional decoding is required, whole or a part of the message needs to be received and stored in advance. Having redundancy greater than Huffman code means that in general we need more bits to code each symbol of the alphabet. Then obviously it would be reasonable to tolerate a short delay in order not to waste our resources. Another motivation is that we don't have any fix-free code for some length vectors like anti-uniform sources where the optimal code length vector is like $L^{C}~=~(1,2,3, \ldots, n-1,n-1)$.

\section{Introduction}
\subsection{Backward Decoding Algorithm}

For backward decoding of Huffman codes the algorithm is obviously the same as the one for decoding the uniquely decodable codes since the backward code of a Huffman code is uniquely decodable. For example, imagine the code $\mathcal{C}$ as :
‎$$\mathcal{C}=\{00‎, ‎10‎, ‎001‎, ‎101‎, ‎011‎, ‎111\}$$‎‏

\noindent The code tree of $\mathcal{C}$ is depicted in Fig.\ref{fig.BDCode1}. The algorithm works with a list of scenarios $\mathcal{L}$, where each scenario, $S_i$, consists of a codeword string and a pointer to one of the nodes in the code tree. At the beginning there is only one scenario in the list that is: $S_1=\{P(Root)\}$. By processing each bit the list is updated and the pointers of each scenario will move their path on the tree. The steps are described bellow:

\begin{figure}[h!]
  \centering
  \includegraphics[width=4cm]{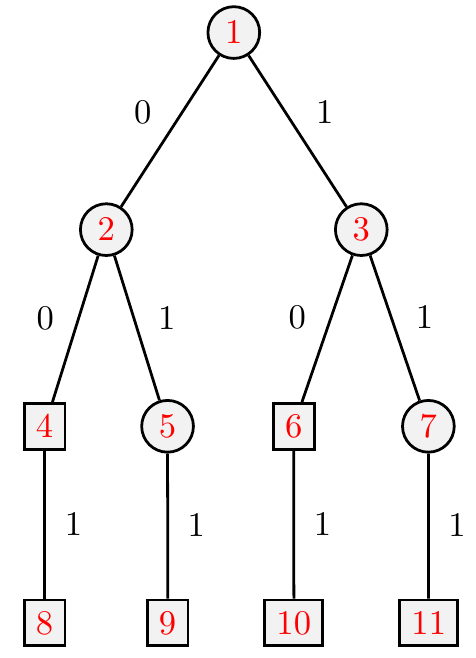}\\
  \caption{\small Code tree of $\mathcal{C}$.}\label{fig.BDCode1}
\end{figure}
\begin{enumerate}
  \item If the pointer comes to a middle node which represents a codeword $c_j$ (square nodes), the pointer of scenario will be updated and a new scenario will be added as : $S_i~=~\{\cdots ,c_j,P(Root)\}$.
  \item If the square node was a leaf, the current scenario will be deleted and the previous mentioned scenario will be added to the list.
  \item If the pointer comes to a null node, the scenario will be deleted.
  \item Just before processing the next bit, first codewords of all the scenarios are to be compared and if the initial codewords of all scenarios were the same, the codeword would be decoded and eliminated from the scenarios. This process goes on until the initial codewords differ.
\end{enumerate}
\subsection{Delay}
\vspace{0mm}
For same code of the previous example, if the text to be decoded is $T=001000 \cdots 00$ then while processing the middle zero bits there is always an ambiguity between two scenarios:

‎\begin{eqnarray*}‎
‎‎&\mathcal{C}=\{
 {c}_{1}: 00‎,
 ‎{c}_{2}: 10, ‎
 ‎{c}_{3}: 001,‎
 ‎{c}_{4}: 101‎,
 ‎{c}_{5}: 011‎,
 ‎{c}_{6}: 111‎\}\\
‎&T=0010000 \cdots 0^{\downarrow}000 \cdots 000 \cdots \\‎
&S_{1}=\{c_{3},c_{1},c_{1},c_{1}‎, ‎\cdots‎, ‎c_{1},C(1)\}\\‎
‎&S_{2}=\{c_{1},c_{2},c_{1},c_{1}‎, ‎\cdots‎, ‎c_{1},C(2)\}‎
\end{eqnarray*}

In general it could be shown that there always exists a bit string in which the ambiguity will remain until we process the whole text \cite{Schutzenberger1966,Capocelli1979}. However, by the average delay approach we will show that, there is almost always a huge difference between equivalent codes \cite{Norwood1967} of a given length vector.

\subsection{List's Size}

We can easily find an upper bound on the list's size.

\begin{lemma}
 The pointers of different scenarios are pointing to nodes from different levels of the code tree.
\end{lemma}\label{lemma1}
\begin{proof}
The representative code of one of two different pointers of different scenarios should be prefix of the other, since they have to be the binary translation of a unit text. If the pointers belong to the same level of tree, this means that these codes should be equal and thus, the pointers are the same which leads to the equivalence of the scenario. Also it is easy to show that same copies of a unique scenario would not be existed. \cite{Fraenkel1990}
\end{proof}

\begin{theorem}
The upper bound for the list's size $|\mathcal{L}|$ is :
\end{theorem}
\begin{equation}
‎|\mathcal{L}‎| \leq \min\{\lfloor\log_{\Phi}\left(\dfrac{\Phi+1}{p_{m}\Phi + p_{m-1}‎}\right)‎ \rfloor ‎, m‎ - ‎1‎‎\}‎‎‎
\end{equation}

where $\Phi=\dfrac{1+\sqrt{5}‎}{2}‎$‎‎‎ and $‎‎‎m=‎‎‎‎‎|\mathcal{A}|‎$ is the size of the alphabet.

\begin{proof}
Using lemma \ref{lemma1} we can easily show that the size of the list is the same as the number of the levels of the code tree. In \cite{Buro1993} by using the compactness of the Huffman codes, it is proven that the maximum length of the longest codeword of a Huffman code is bounded to $\min\{\lfloor\log_{\Phi}(\dfrac{\Phi+1}{p_{m}\Phi + p_{m-1}‎})‎ \rfloor ‎, m‎~-~‎1‎‎\}‎‎‎$, which is also a bound for the number of the code tree's levels.
\end{proof}

\subsection{Properties}

Some important properties of the algorithm are to mention in order to be used in next sections.

\textbf{\emph{Property 1 }}: In each iteration of bit processing the number of scenarios to be added to the list is at most one.

This is a good remark which guarantees that the need for the memory of this algorithm is bounded and self controlled.

\textbf{\emph{Property 2 }}: Non of the pointers point at the leafs of the tree. Recalling the fact that the leafs are definitely square nodes, and also the second step of the algorithm, it is obvious.

\textbf{\emph{Property 3 }}: According to the first step, it is clear that if the pointer comes to a middle square node, two scenarios with same initial codewords can be found in the list. This \emph{Twin Scenarios} property will help us to propose a new search algorithm in order to find a minimum delay code.

\begin{figure}[h]
  \centering
  \includegraphics[height=5cm]{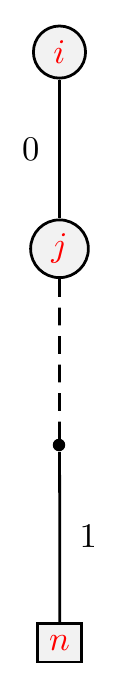}\\
  \caption{\small$i-n$ branch.}\label{fig:NegDelay}\vspace{-5mm}
\end{figure}

\textbf{\emph{Property 4 }}: Imagine There is only one scenario in the list, that is $S_1=\{P(i)\}$. If the branch $i-n$ depiceted in Fig.\ref{fig:NegDelay} does not have any sub branches and middle square nodes, there will be a minus delay of $-k$, where $k$ is the length of the branch. It is obvious that in the case that there is no other scenarios in the list, the only possible codeword is the $c_n$. This property seems interesting, however; we will show that for most of the time this feature wont help us to reduce the average delay.

\section{Delay Analysis and Theoretical Calculations}

\subsection{Delay Definition}

\begin{definition}
First we define the Maximum Backward Decoding Delay (M-BDD): {\it"number of bits which the decoder has to process more than the last bit of the first symbol decoded in each section of decision making."}
\end{definition}\vspace{-5mm}

\begin{figure}[!h]\
\centering
\includegraphics[width=4.2cm]{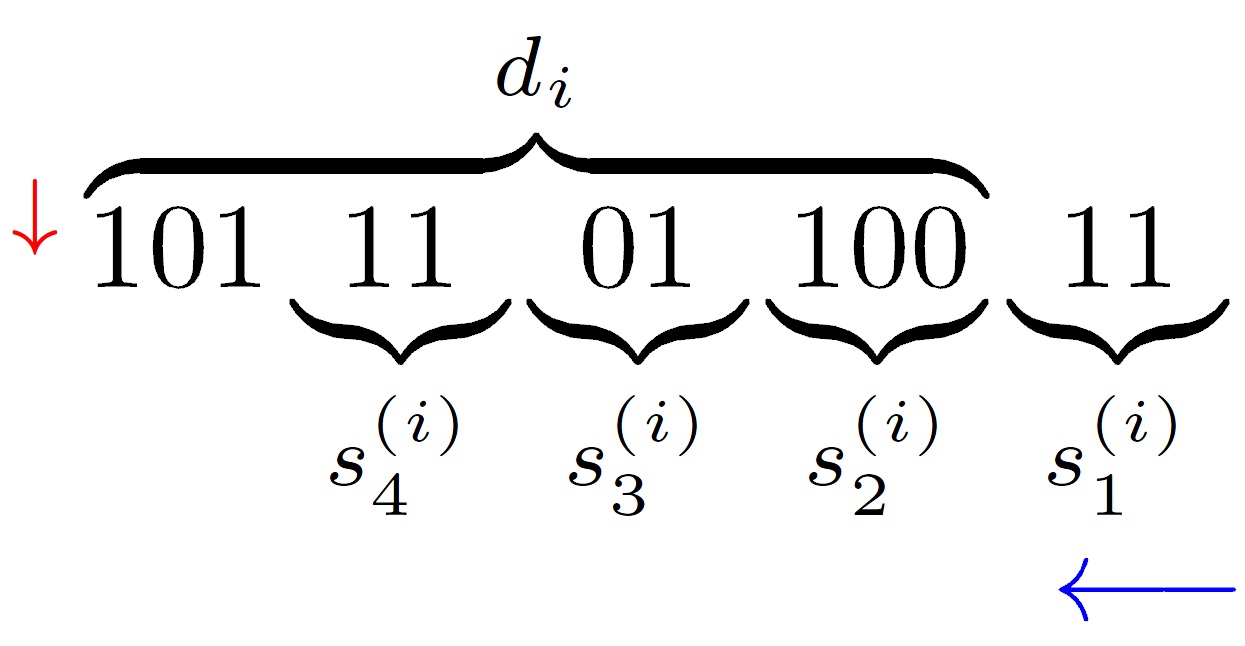}\\
\caption{\small M-BDD $d_i$ of a given binary text.}
\label{fig:di}‎‎\vspace{-5mm}
‎\end{figure}

\begin{example}
For better understanding consider this Example. Imagine the code table in the form below:\vspace{-5mm}
\end{example}
\begin{table}[h]
\caption{\small Code book of the Example.}
\centerline{
\begin{tabular}{cc}
  \toprule
  Symbol & Codeword \\
  \midrule
  a & 0 \\
  b & 10 \\
  c & 11 \\
  \bottomrule
\end{tabular}}
\end{table}

Two examples of arbitrary strings of symbols with the corresponding coded strings are shown with their relative Maximum Backward Decoding Delays:

\begin{table}[h]
\centerline{
\begin{tabular}{l l}
 Symbol string:& a c c c c c b \\
Bit String: & $\underbrace{01111111111}_{M-BDD}10$ \\
 Symbol string:& a c c c c c a \\
Bit String: & $\underbrace{01111111111}_{M-BDD}0$ \\
\end{tabular}}
\end{table}

\begin{definition}
The average M-BDD of a code is defined as: {\it"The expectation value of each probable Maximum Backward Decoding Delay."} If each probable scenario of decision making has the probability $q_{i}$ and the correspondence delay is $d_{i}$, the average M-BDD will be calculate by Eq.~(\ref{AvgMBDD}):
\begin{equation}\label{AvgMBDD}
  \overline{D}=\sum_{i}{q_{i}d_{i}}
\end{equation}
\end{definition}
Therefore, a code with minimum M-BDD has the best performance in the worst case.

\subsection{Dependance of Distribution on The Decision Making Index}

\begin{figure}[!h]
\centering
\includegraphics[width=8.6cm]{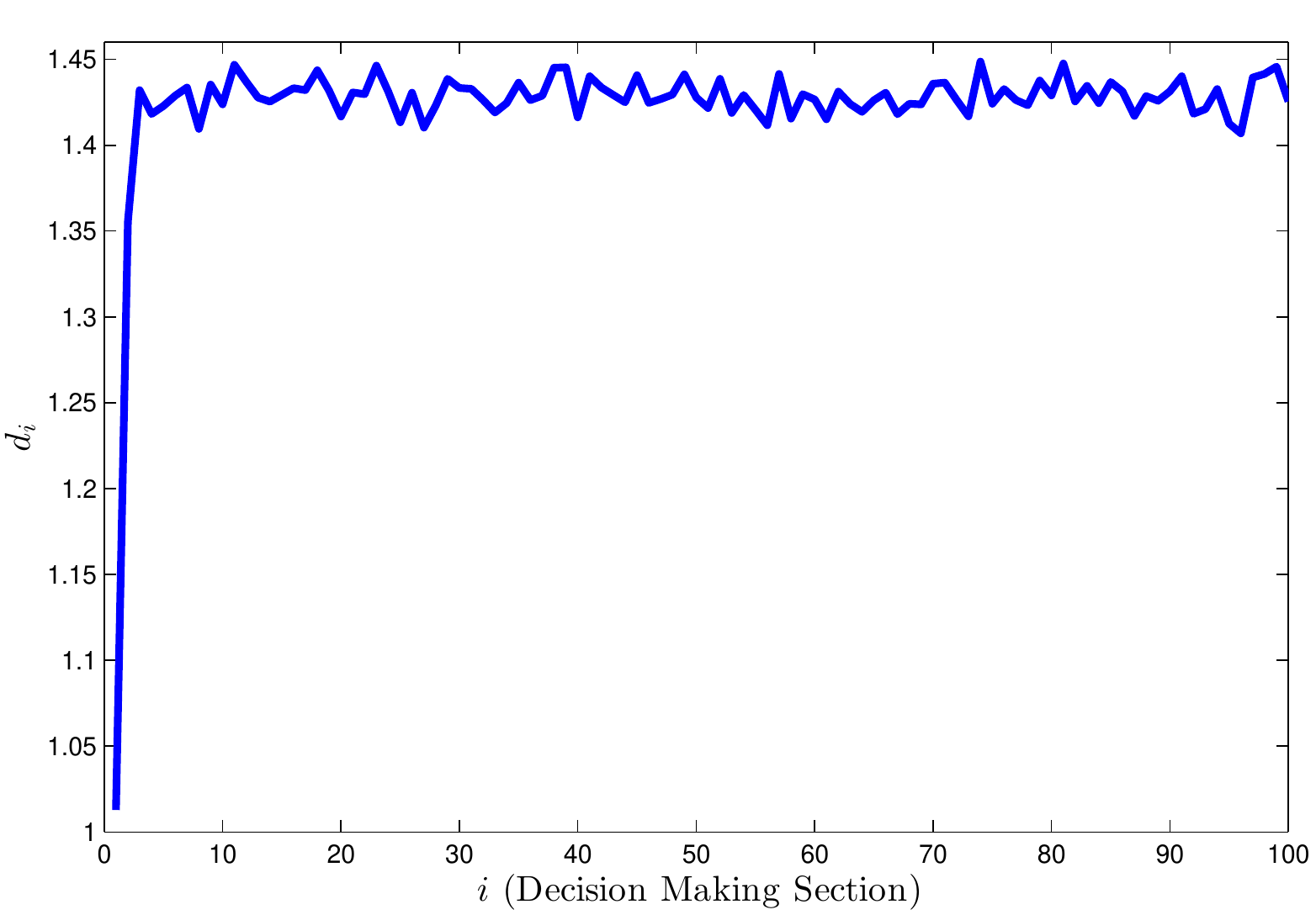}\\‎
\caption{\small The average M-BDD plot vs. decision making ‎index.} ‎‎
\label{fig:dii}‎‎\vspace{-5mm}
‎\end{figure}

Decision making point is the point in which the algorithm decodes one or more codewords. Imagine the anti-uniform code $\mathcal{C}=\{0‎, ‎11‎, ‎100‎, ‎101\}$. For $20000$ bit strings of $200$ symbols with the source distribution of $‎‎P=(\frac{1}{2},\frac{1}{4},\frac{1}{8},\frac{1}{8})$
\noindent the average M-BDD plot for each index of decision making point is depicted in Fig. \ref{fig:dii}.\vspace{-5mm}

This shows how decoding algorithm changes the relevant distribution of probable delays at each decision making point.
\subsection{Theoretical Analysis for Canonical Anti-Uniform Codes}

The anti-uniform source is a source which has Huffman length vector in the form $L^{C}=(1,2,3, \ldots, n-1,n-1)$.
\begin{figure}[!h]
\centering
\begin{minipage}[r]{0.18\textwidth}
\centering
\centering
\includegraphics[width=3cm]{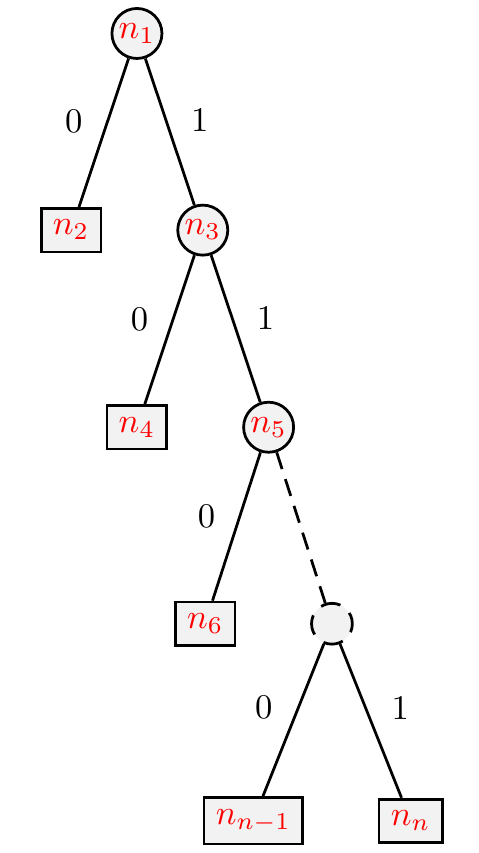}\\‎

\end{minipage}
\begin{minipage}[l]{0.23\textwidth}
\centering
{\renewcommand{\arraystretch}{0.8}
\begin{tabular}{cc}

  \toprule
  \small{Symbol} & \small{Codeword} \\
  \midrule
  $s_1$ & 0 \\
  $s_2$ & 10 \\
  $s_3$ & 110 \\
  $s_4$ & 1110 \\
  $s_5$ & 11110 \\
  $s_6$ & 111110 \\
  $\vdots$ & $\vdots$ \\
  $s_{n-1}$ & $\underbrace{11\ldots1}_{n-2}0$ \\
  $s_{n}$ & $\underbrace{11\ldots11}_{n-1}$ \\
  \bottomrule
\end{tabular}}
\end{minipage}
\caption{\small Canonical Anti-Uniform Source Code.}\label{cananti}\vspace{-5mm}
\end{figure}
You can think of a source with a probability distribution of the form~: $P=(\frac{1}{2},\frac{1}{4} ,\frac{1}{8}, \ldots , \frac{1}{2^{n-1}} ,\frac{1}{2^{n-1}})$. For these sources obviously there is no fix-free codes. A Canonical Anti-Uniform Huffman Code has a code book as shown in Fig.\ref{cananti}.

\begin{figure}[h]
  \centering
  \includegraphics[width=8.6cm]{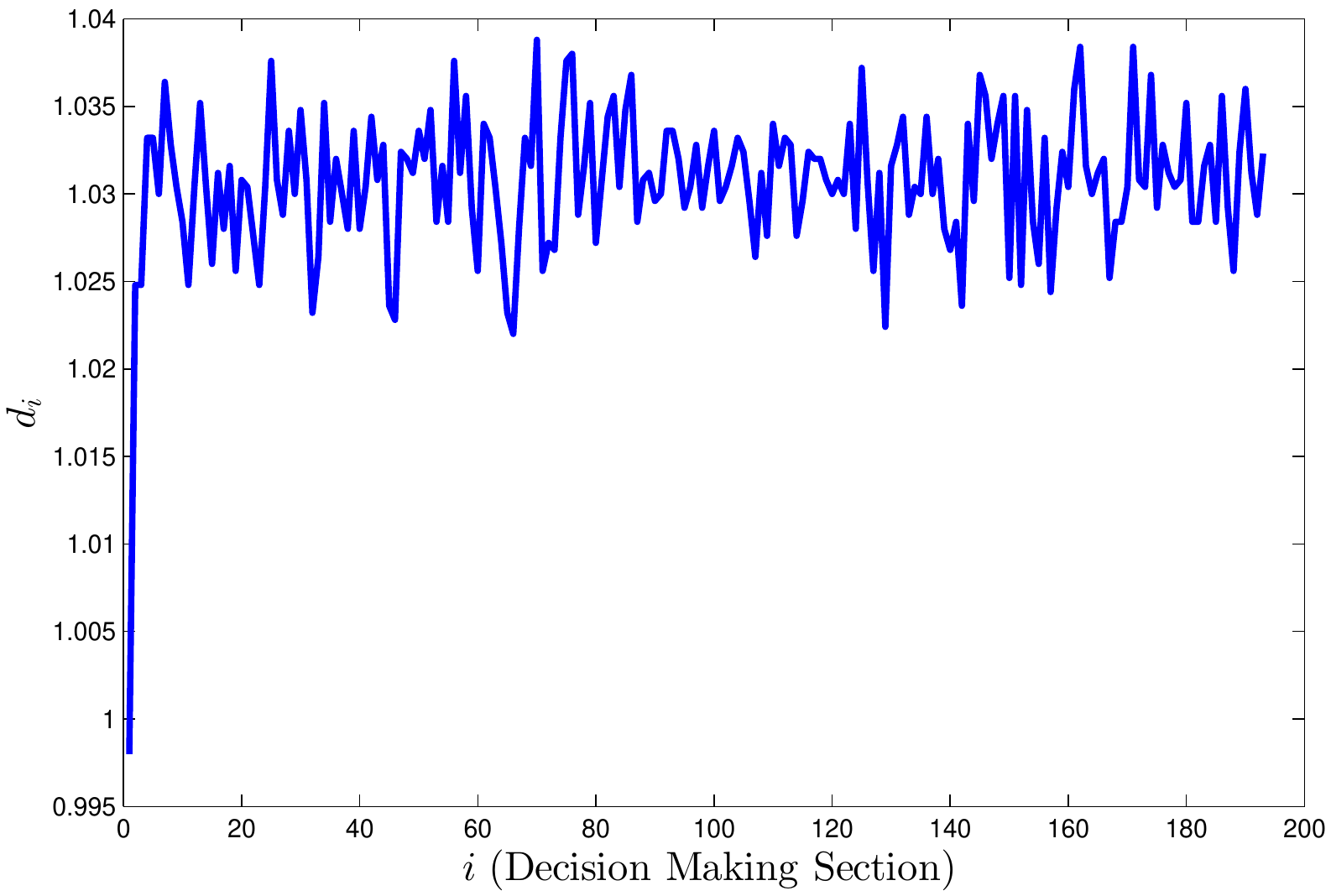}\\
  \caption{\small Plot for the canonical anti-uniform code with $|\mathcal{A}|=9$.}\label{fig:dii9}\vspace{-5mm}
\end{figure}

Based on the simulations the canonical codes has the least M-BDD among all anti-uniform Huffman codes. For instance, for the alphabet size of $9$ the average M-BDD of the canonical anti-uniform code is $1.03$ bits. The average M-BDD plot of this code with same conditions of Fig.\ref{fig:dii} is as shown in Fig.\ref{fig:dii9}.

\begin{figure}[bh]
  \centering
  \includegraphics[height=5cm]{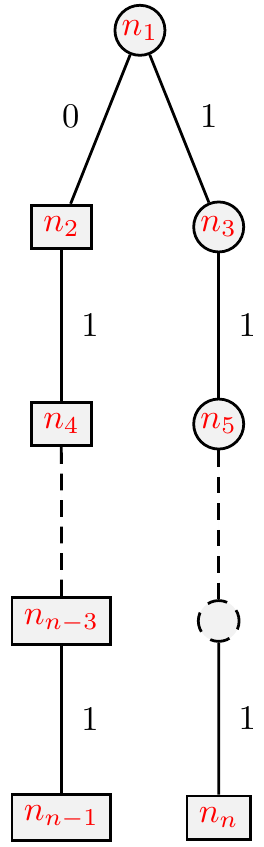}\\
  \caption{\small Canonical Anti-Uniform Backward Code Tree.}\label{fig:canti}\vspace{-3mm}
\end{figure}

In order to theoretically calculate their backward decoding delay notice the backward code tree of a typical canonical anti-uniform code which is depicted in Fig. \ref{fig:canti}.

As shown in the Fig.\ref{fig:canti} it is clear that according to the fourth property of decoding algorithm, we expect negative delays in the initial iterations and that's why the plot of Fig.\ref{fig:dii9} has smaller values in the initial indexes. According to the distribution changes in following decision making sections, we can easily ignore the initial minus delays and just assume that the text does not end in the last symbol. This means at the end of the bit string we have a zero following by some ones. Then easily by checking the table \ref{tab:canti} one can calculate the average backward decoding delay.

\begin{table}[h!]
\centering
\caption{\small Different Sessions of the delay calculation for canonical anti-uniform code.}\label{tab:canti}
\begin{tabular}{cccc}
  \toprule
  Symbols & Bit String & $d_{i,j}$ & $p_{i,j}$ \\
  \midrule
  $s_k \underbrace{s_n \ldots s_n}_{j} s_1$ & $\ldots0\underbrace{1\ldots1}_{j(n-1)}0$ & $j(n-1)+1$ & $p_{1}(p_n)^{j}$ \\
  $s_k \underbrace{s_n \ldots s_n}_{j} s_2$ & $\ldots0\underbrace{1\ldots1}_{j(n-1)}10$ & $j(n-1)+1$ & $p_{2}(p_n)^{j}$ \\
  $s_k \underbrace{s_n \ldots s_n}_{j} s_3$ & $\ldots0\underbrace{1\ldots1}_{j(n-1)}110$ & $j(n-1)+1$ & $p_{3}(p_n)^{j}$ \\
  $\vdots$ & $\vdots$ & $\vdots$ & $\vdots$ \\
  $s_k \underbrace{s_n \ldots s_n}_{j} s_{n-1}$ & $\ldots0\underbrace{1\ldots1}_{j(n-1)}\underbrace{1\ldots1}_{n-2}0$ & $j(n-1)+1$ & $p_{n-1}(p_n)^{j}$ \\
  \bottomrule
\end{tabular}
\vspace*{2mm}
\centering \tiny \\ Note: $k\neq n$ \& $j=0,1,\ldots$ \vspace{-3mm}
\end{table}

Thus we will have:
\begin{eqnarray*}
  \overline{D} &=& \sum_{i,j}{d_{i,j}p_{i,j}} \\
   &=& \sum_{i=1}^{n-1}{p_i}\sum_{j=0}^{\infty}{(p_{n})^{j}(j(n-1)+1)} \\
   &=& (1-p_{n})\sum_{j=0}^{\infty}{(p_{n})^{j}(j(n-1)+1)}
\end{eqnarray*}
For a diadic source distribution we have $p_{i}=\frac{1}{2^{l_i}}$, so a canonical anti-uniform Huffman code with diadic distribution $P=(\frac{1}{2},\frac{1}{4},\ldots,\frac{1}{2^{n-1}},\frac{1}{2^{n-1}})$, has the average maximum backward decoding delay as:
\begin{equation}
  \overline{D}=(1-\frac{1}{2^{n-1}})\sum_{j=0}^{\infty}{\left(\frac{1}{2^{n-1}}\right)^{j}(j(n-1)+1)})
\end{equation}
By straight forward calculation we can simplify the above into:
\begin{equation}\label{eq:dbar}
  \overline{D}=\frac{n-1}{2^{n-1}-1}+1
\end{equation}
Easily we can see that for a sufficiently large source we have:
\begin{equation}
  \lim_{n\rightarrow\infty}\overline{D}=\lim_{n\rightarrow\infty} \frac{n-1}{2^{n-1}-1}+1 = 1 ~~bit
\end{equation}

For sources with different alphabet sizes the results of comparisons between equation (\ref{eq:dbar}) and simulations could be found in Table \ref{tab:anti}.
\begin{table}[!h]
\centering
\caption{\small Comparison of Equation \ref{eq:dbar} with Simulations.}\label{tab:anti}
\begin{tabular}{|c|c|c||c|c|c|}
      \hline
      n & $\overline{D}_{Theory}$ & $\overline{D}_{Simulation}$ & n & $\overline{D}_{Theory}$ & $\overline{D}_{Simulation}$ \\ \hline \hline
      3 & 1.66 & 1.66        &  8 & 1.0551 & 1.0535                  \\ \hline
      4 & 1.428 & 1.414     ‎& 9 & 1.0313 & 1.0369                 \\ \hline
      5 & 1.266 & 1.259     &  10 & 1.0176 & 1.0179                 \\ \hline
      6 & 1.1612 & 1.1730 &   11 & 1.010 & 1.010                 \\ \hline
      7 & 1.0952 & 1.0861  &   12 & 1.005 & 1.005          \\ \hline
    \end{tabular}\vspace{-5mm}
\end{table}

\section{In Search of Minimum Delay Code}

Here we want to define a search algorithm so that it can find the minimum delay code. For this purpose, we will define some new measures in order to compare equivalent Huffman Codes in the sense of their backward decoding delay. First, we remark some standard notions regarding the binary parsing trees. \cite{Knu98}

Using the backward code tree of a Huffman code we make the $\mathcal{E}$ tree by including the null nodes. Thus, here by the tree has three types of nodes:
\begin{enumerate}
  \item $\mathcal{V}$ : The middle square nodes (the codeword nodes).
  \item $\mathcal{X}$ : The null nodes.
  \item $\mathcal{U}$ : Other nodes.
\end{enumerate}

\begin{figure}[h!]
  \centering
  \includegraphics[width=7.5cm]{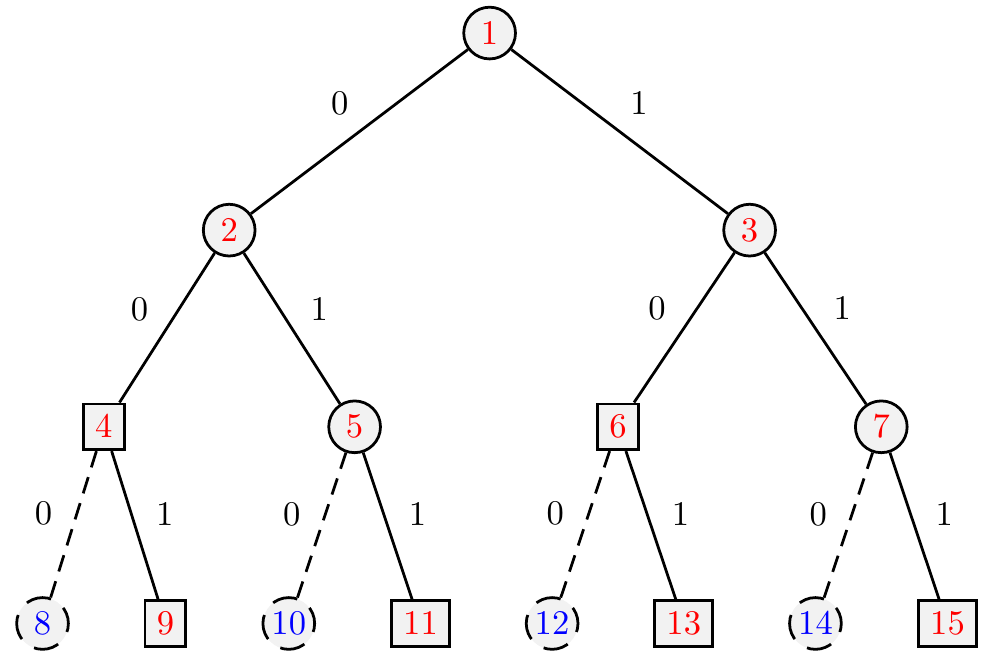}\\
  \caption{\small Relevant $\mathcal{E}$ tree of code tree of Fig.\ref{fig.BDCode1}.}\label{fig:Etree}\vspace{-5mm}
\end{figure}

According to the third property of the decoding algorithm, if a pointer comes to a middle square node, in the list we will have two scenarios with same initial codewords. As the number of scenarios in the list is bounded, if the production of these type of twin scenarios occurs more often the chance to have a list of scenarios filed with all scenarios of same initial codewords will increase. Therefore, based on the last step of the algorithm we can have decision making points, in which at least one codeword has been decoded, closer to each other.

The $m^+$ measure will represent the chance of pointers to point to middle square nodes. As we can assume that the probability of pointing to each node is close to the $2^{-level(n)}$, where $level(n)$ indicates the level of each node, which here is the same as the square node's codeword length, $l_i$.

\begin{equation}
m^+ = \sum_{n \in \mathcal{V}} 2^{-level(n)}
\end{equation}

Also as each scenario will be updated in the next iteration, algorithm may produce twin scenarios, terminate the scenario or just updating its pointer. The value of each change is different, since we are interested to have twin scenarios to be produce in each iteration of bit processing and to have them remained in the list. A conventional way is to model these changes' values by the $\eta$ function:

‎‎\begin{eqnarray}
\eta ‎(n_i) = \left\{
\begin{array}{cr}
+1 & n_i \in \mathcal{V} \\
0 & n_i  \in \mathcal{U} \\
-1 & n_i \in \mathcal{X}
\end{array} \right\}
\end{eqnarray}

If $n_1$ and $n_2$ are the two children of the node $n$, $\{n_1,n_2\}~=~Children(n)$ , then by defining the set $\mathcal{Q}$ as
$$\mathcal{Q}~=~Children(\mathcal{V})\cup Children(Root)$$
we can define the $m^{\delta}$ measure as follows:

\begin{equation}
m^\delta = \sum _{n \in \mathcal{Q}‎}\eta(n)\cdot ‎‎‎2^{-level(n)}
\end{equation}‎

This measure will indicate the chance of twin scenarios to remain in the list in future iterations. Finally, as we expect the minimum delay code to has both high $m$ measures, the $M$ measure is defined as the sum of these factors.

\begin{equation}
M=m^+ + m^\delta
\end{equation}‎‎

Now our search algorithm would be an easy comparison between the $M$ factor of all Huffman codes for a given length vector. Our results of simulations for all source with alphabet sizes of $7$, $8$, and $9$ are shown in Tables bellow. $|\mathcal{C}|$ is the number of all equivalent Huffman codes for a given length vector. The $\Delta D$ column shows the error of our proposed search algorithm, which indeed shows good performance.

\setlength{\tabcolsep}{3pt}
\begin{table}[h]
  \centering
  \caption{\small Results for all sources with $|\mathcal{A}|=7$.}
    \begin{tabular}{cccccc}
    \toprule
          &       & \multicolumn{2}{c}{Simulations} & \multicolumn{2}{c}{Search Algorithm}  \\
    \midrule
    $L$  & $|\mathcal{C}|$   & Min $\overline{D}$ & Max $\overline{D}$ & $\overline{D}$  & $\Delta D$ \\ \midrule
    (2,2,2,3,4,5,5) & 16    & 6.728 & 35.624 & 7.483 & 0.755 \\
    (2,2,2,4,4,4,4) & 4     & 1.608 & 1.608 & 1.608 & 0 \\
    (2,2,3,3,3,4,4) & 24    & 4.450 & 40.653 & 4.450 & 0 \\
    (2,3,3,3,3,3,3) & 4     & 12.956 & 39.839 & 12.956 & 0 \\
    (1,2,3,4,5,6,6) & 32    & 1.094 & 39.466 & 1.094 & 0 \\
    (1,2,3,5,5,5,5) & 8     & 3.250 & 13.566 & 3.250 & 0 \\
    (1,2,4,4,4,5,5) & 16    & 3.257 & 10.703 & 3.460 & 0.203 \\
    (1,3,3,3,4,5,5) & 16    & 3.331 & 13.123 & 4.198 & 0.867 \\
    (1,3,3,4,4,4,4) & 12    & 4.036 & 7.838 & 4.036 & 0 \\ \midrule
    \multicolumn{2}{c}{Average}   & 4.523 & 22.491 &       & 0.203 \\
    \bottomrule
    \end{tabular}%
  \label{tab:addlabel}%
\end{table}%

\setlength{\tabcolsep}{3pt}
\begin{table}[h]
  \centering
  \caption{\small Results for all sources with $|\mathcal{A}|=8$.}
    \begin{tabular}{cccccc}
    \toprule
          &       & \multicolumn{2}{c}{Simulations} & \multicolumn{2}{c}{Search Algorithm}  \\
    \midrule
    $L$  & $|\mathcal{C}|$   & Min $\overline{D}$ & Max $\overline{D}$ & $\overline{D}$  & $\Delta D$ \\ \midrule
    (2,2,2,3,4,5,6,6) & 32    & 7.488 & 42.357 & 7.488 & 0 \\
    (2,2,2,3,5,5,5,5) & 8     & 8.211 & 28.391 & 8.952 & 0.741 \\
    (2,2,2,4,4,4,5,5) & 16    & 18.819 & 67.086 & 18.819 & 0 \\
    (2,2,3,3,3,4,5,5) & 48    & 6.079 & 25.644 & 6.189 & 0.110 \\
    (2,2,3,3,4,4,4,4) & 36    & 6.086 & 36.040 & 7.939 & 1.852 \\
    (2,3,3,3,3,3,4,4) & 24    & 6.774 & 55.138 & 6.774 & 0 \\
    (3,3,3,3,3,3,3,3) & 1     & 13.063 & 55.138 & 13.063 & 0 \\
    (1,2,3,4,5,6,7,7) & 64    & 1.057 & 60.347 & 1.057 & 0 \\
    (1,2,3,4,6,6,6,6) & 16    & 3.147 & 25.035 & 3.147 & 0 \\
    (1,2,3,5,5,5,6,6) & 32    & 3.097 & 20.024 & 3.111 & 0.014 \\
    (1,2,4,4,4,5,6,6) & 32    & 3.274 & 13.746 & 3.403 & 0.129 \\
    (1,2,4,4,5,5,5,5) & 24    & 3.229 & 14.043 & 3.354 & 0.125 \\
    (1,3,3,3,4,5,6,6) & 32    & 3.784 & 19.109 & 4.181 & 0.397 \\
    (1,3,3,3,5,5,5,5) & 8     & 4.267 & 13.254 & 4.267 & 0 \\
    (1,3,3,4,4,4,5,5) & 48    & 3.641 & 12.113 & 3.871 & 0.230 \\
    (1,3,4,4,4,4,4,4) & 8     & 4.356 & 9.025 & 4.356 & 0 \\ \midrule
   \multicolumn{2}{c}{Average} & 5.630 & 31.031 &       & 0.225 \\
    \bottomrule
    \end{tabular}%
  \label{tab:addlabel}%
\end{table}%

\setlength{\tabcolsep}{3pt}
\begin{table}[h]
  \centering
  \caption{\small Results for all sources with $|\mathcal{A}|=9$.}
  {\renewcommand{\arraystretch}{1}
    \begin{tabular}{cccccccc}
    \toprule
          &       & \multicolumn{2}{c}{Simulations} & \multicolumn{2}{c}{Search Algorithm}  \\
    \midrule
    $L$  & $|\mathcal{C}|$   & Min $\overline{D}$ & Max $\overline{D}$ & $\overline{D}$  & $\Delta D$ \\ \midrule
    (2,2,2,3,4,5,6,7,7) & 64    & 7.222 & 44.570 & 7.920 & 0.698 \\
    (2,2,2,3,4,6,6,6,6) & 16    & 11.044 & 40.799 & 11.075 & 0.031 \\
    (2,2,2,3,5,5,5,6,6) & 32    & 5.982 & 37.447 & 5.982 & 0 \\
    (2,2,2,4,4,4,5,6,6) & 32    & 34.993 & 112.367 & 34.993 & 0 \\
    (2,2,2,4,4,5,5,5,5) & 24    & 10.576 & 34.471 & 34.471 & 23.895 \\
    (2,2,3,3,3,4,5,6,6) & 96    & 5.248 & 54.348 & 6.117 & 0.869 \\
    (2,2,3,3,3,5,5,5,5) & 24    & 9.790 & 17.570 & 9.790 & 0 \\
    (2,2,3,3,4,4,4,5,5) & 144   & 4.276 & 64.249 & 7.465 & 3.189 \\
    (2,2,3,4,4,4,4,4,4) & 24    & 12.030 & 41.574 & 12.218 & 0.187 \\
    (2,3,3,3,3,3,4,5,5) & 48    & 7.445 & 127.587 & 7.445 & 0 \\
    (2,3,3,3,3,4,4,4,4) & 60    & 0 & 33.112 & 6.020 & 6.020 \\
    (3,3,3,3,3,3,3,4,4) & 8     & 33.262 & 79.019 & 33.471 & 0.209 \\
    (1,2,3,4,5,6,7,8,8) & 128   & 1.027 & 105.722 & 1.027 & 0 \\
    (1,2,3,4,5,7,7,7,7) & 32    & 3.056 & 44.114 & 3.056 & 0 \\
    (1,2,3,4,6,6,6,7,7) & 64    & 3.024 & 35.558 & 3.043 & 0.018 \\
    (1,2,3,5,5,5,6,7,7) & 64    & 3.051 & 25.370 & 3.157 & 0.106 \\
    (1,2,3,5,5,6,6,6,6) & 48    & 3.061 & 25.011 & 3.126 & 0.064 \\
    (1,2,4,4,4,5,6,7,7) & 64    & 2.984 & 15.837 & 3.362 & 0.377 \\
    (1,2,4,4,4,6,6,6,6) & 16    & 3.460 & 12.112 & 3.460 & 0 \\
    (1,2,4,4,5,5,5,6,6) & 96    & 3.141 & 19.986 & 3.275 & 0.135 \\
    (1,2,4,5,5,5,5,5,5) & 16    & 3.274 & 13.958 & 3.274 & 0 \\
    (1,3,3,3,4,5,6,7,7) & 64    & 3.735 & 23.543 & 4.155 & 0.420 \\
    (1,3,3,3,4,6,6,6,6) & 16    & 4.200 & 18.584 & 4.200 & 0 \\
    (1,3,3,3,5,5,5,6,6) & 32    & 3.729 & 17.673 & 4.479 & 0.749 \\
    (1,3,3,4,4,4,5,6,6) & 96    & 3.281 & 16.307 & 3.961 & 0.680 \\
    (1,3,3,4,4,5,5,5,5) & 72    & 3.640 & 14.973 & 3.928 & 0.287 \\
    (1,3,4,4,4,4,4,5,5) & 48    & 4.261 & 14.806 & 4.411 & 0.150 \\
    (1,4,4,4,4,4,4,4,4) & 2     & 8.763 & 8.763 & 8.763 & 0 \\ \midrule
    \multicolumn{2}{c}{Average}  & 7.127 & 39.265 &       & 1.360 \\
    \bottomrule
    \end{tabular}}
  \label{tab:addlabel}
\end{table}
\section{Conclusion}

In this paper we studied the problem of Bidirectional Decoding of Huffman Codes. However, Huffman Codes are the only optimal codes in the sense of redundancy \cite{Huffman1952}, for the backward decoding procedure of these codes we will face to the problem of decoding delay. Despite the traditional belief that there wont be any noticeable difference between equivalent codes of a given length vector \cite{Fraenkel1990}, we showed that for almost all the cases there exists a huge difference. The common fix-free codes that have been considered for the fast bidirectional decoding applications may increase the redundancy of the code, which is the main issue for those applications where we encounter storage limitations. Therefore, for the applications where both bidirectional decoding and redundancy limitations are of concern finding the minimum delay Huffman codes is crucial. For the special case of anti-uniform sources we showed that the average backward decoding delay is near one. Also by considering some properties of the decoding algorithm, we proposed a search protocol which will find a Huffman code with sufficiently short delay for a given length vector $L$.

For future research we should try to find a precise proof to show the exact relation of delay and decoding complexity which is the main factor to compare the speed of decoding algorithms. The other subject is to figure out new rules to help us in order to use and take into account other codes (for example other Huffman codes) for a given length vector to balance the tradeoff between redundancy and delay costs.

\bibliographystyle{ieeetr}
\bibliography{Proj1}

\end{document}